\documentclass[12pt]{article}

\usepackage{amsmath,amsfonts,amssymb,color,graphicx,overpic,}
\usepackage{slashed}
\usepackage{epsfig}
\usepackage[utf8]{inputenc}
\usepackage{amsmath}
\usepackage{ wasysym }
\usepackage{graphicx}
\usepackage[english]{babel}
\usepackage{graphic x}
\usepackage{relsize}
\usepackage{cite}

\begin{document}

\begin{titlepage}
\rightline{April 2018}
\vskip 1.9cm
\centerline{\large \bf
DAMA annual modulation from electron recoils}

\vskip 2.1cm
\centerline{R. Foot\footnote{
E-mail address: rfoot@unimelb.edu.au}}

\vskip 0.4cm
\centerline{\it ARC Centre of Excellence for Particle Physics at the Terascale,}
\centerline{\it School of Physics, The University of Sydney, NSW 2006, Australia}
\vskip 0.1cm
\centerline{and}
\vskip 0.1cm
\centerline{\it ARC Centre of Excellence for Particle Physics at the Terascale,}
\centerline{\it School of Physics, The University of Melbourne, VIC 3010, Australia
}

\vskip 2.0cm
\noindent
Plasma dark matter, which arises in dissipative dark matter models, 
can give rise to large annual modulation signals from keV electron recoils.
Previous work has argued that the DAMA annual modulation signal 
could have an explanation within such a scenario. 
However, detailed predictions are difficult due
to the inherent complexities involved in modelling the halo plasma interactions with Earth-bound dark matter.
Here, we consider a simple phenomenological model for the dark matter velocity function 
relevant for direct detection experiments, and compare the resulting electron scattering rate with the new DAMA/LIBRA phase 2 data.
We also consider the constraints from other experiments, including XENON100 and DarkSide-50.

\end{titlepage}

\section{Introduction}

The DAMA collaboration have
observed an annually modulating scintillation rate in a NaI target for over a decade, with properties broadly consistent
with a dark matter signal \cite{dama1,dama2a,dama2b,dama3,dama4}.
An interpretation of the DAMA annual modulation in terms of nuclear recoils appears to be excluded
by many other experiments, including XENON1T \cite{xenon1T}, DarkSide \cite{darkside1}, PANDA \cite{panda}, 
LUX \cite{lux}, CRESST \cite{cresst}, CDMS \cite{cdms}, XMASS \cite{xmass}, PICO \cite{pico}, etc.
However the DAMA experiment is sensitive to both electron and nuclear recoils, and if the annual modulation is due to 
electron recoils then it is far more weakly constrained.
% by the other experiments.
This suggests that dark matter
scattering off electrons 
is the more likely explanation, if the DAMA signal is indeed due to dark matter interactions.

Electron recoils with keV energy scale can arise from dark matter scattering 
if there are dark matter particles of mass $\sim$ MeV with kinetic energy in the keV 
range \cite{foot14,JC1}. 
Such particles occur in dissipative dark matter models, where 
the dark matter halo in the Milky Way takes the form of a dark plasma.\footnote{For completeness, we note here
that electron recoils can also arise in WIMP models, and have been discussed in the context of the DAMA experiment in \cite{damaelectron,zupan,roberts}.
In such models,
keV electron recoils result from GeV mass-scale WIMP dark matter interactions with
inner shell atomic electrons. By contrast, the plasma dark matter case has dark matter in the MeV mass range but has rather large kinetic energy
arising from energy equipartition of the light and heavy plasma components. In the plasma case, keV energy electron recoils
readily occur, and can result predominately from dark electron scattering off loosely bound (outer shell) atomic electrons.}
Specifically,  if dark matter consists of dark electrons and dark protons coupled together via
a massless dark photon, e.g. \cite{sunny}, then
energy equipartition implies that the dark electrons and dark protons have the same
temperature and the same mean energy. 
The halo temperature is set by the mean particle mass, $\bar m \equiv \sum n_i m_i/\sum n_i$,
and is estimated to be:
\begin{eqnarray}
T \approx \frac{\bar m v_{\rm rot}^2}{2}
\ .
\label{1}
\end{eqnarray}
Here, $v_{\rm rot}$ is the asymptotic rotational velocity, which for the
Milky Way galaxy is $v_{\rm rot} \approx 220$ km/s.
The above temperature relation indicates that the
kinetic energy of dark electrons with a mass of order a MeV can be in the keV range
provided that $\bar m$ is of order a few GeV or so.

In such models the kinetic mixing interaction \cite{holdom,fhe} is introduced to achieve consistent halo dynamics \cite{sph,review,sunny}. 
The dark halo is dissipative and cools, and with kinetic mixing around $\epsilon \sim 10^{-9}-10^{-10}$ 
type II supernovae can be transformed into powerful dark sector heat sources. 
It has been argued in a number of papers, most recently in \cite{foot18a,foot18,sunny2},
that dissipative dark matter with heating sourced from supernovae
can lead to a successful framework in which to understand small scale structure issues.
Importantly,
this picture can be tested in direct detection experiments as the
kinetic mixing interaction also allows for  observable dark matter scattering off ordinary particles, with
the most favourable detection channels being
dark electron - electron scattering, and dark proton - nuclei scattering, e.g. \cite{JC2}.

Kinetically mixed mirror dark matter provides a theoretically constrained dark plasma model \cite{flv,review}.
The mean halo mass is around a GeV and the 
temperature of the dark halo is estimated to be around $T \approx 0.5$ keV.
Furthermore, the dark electron is the mirror electron, a particle hypothesized to have exactly
the same mass as the electron.
The analysis in this paper is applicable to the mirror dark matter 
case, but also relevant to more generic plasma dark matter models.

Mirror dark matter, and related models, have a number of nontrivial aspects. 
Of particular concern for direct detection experiments is the interaction between the halo wind and
captured dark matter within the Earth.
The captured dark matter provides an obstacle to the halo wind, and can strongly
influence halo dark matter properties (density and distribution) in the Earth's vicinity.
The effects of this `dark sphere of influence' can be explored
with MHD equations, and the temperature and density distributions of the halo dark matter
in the Earth's vicinity studied.
An investigation along these lines \cite{JC1} found that
the annual modulation signal can be greatly enhanced, even a modulation
amplitude near maximal (i.e. of order 1) is possible.
In addition, diurnal variation is also expected to be significant, and can in fact be maximal
for a detector located in the southern hemisphere \cite{foot12,footsunny,JC1}.

The enhanced annual modulation can help reconcile the positive DAMA annual modulation signal 
with the results reported
by XENON100 \cite{xene1,xene2}, as well as stringent constraints from the DarkSide-50 experiment \cite{darkside}. 
Detailed predictions, though, are difficult due
to the inherent complexities involved in modelling the halo plasma interactions with Earth-bound dark matter.
Here, we consider a simple phenomenological model for the dark matter velocity function relevant to direct
detection experiments,
and confront the model with the available data.
It turns out that the DAMA experiment and the DarkSide-50 experiment are the most sensitive probes of 
the electron scattering signal expected within this model, and in fact a self consistent picture
emerges. 

\section{The mirror dark matter model}

The mirror dark matter model assumes that dark matter arises from a hidden sector which is an exact copy of
the standard model. That is, the
Lagrangian describing fundamental physics is
\begin{eqnarray}
{\cal L} = {\cal L}_{\rm SM}(e,u,d,\gamma,...) + {\cal L}_{\rm SM}(e',u',d',\gamma',...)
+ {\cal L}_{\rm mix}
\ .
\label{yyy6}
\end{eqnarray}
The model features an exact unbroken $Z_2$ mirror symmetry, which can be interpreted as space-time parity if the
chirality of the hidden sector  fermions is flipped \cite{flv}.
The mirror sector particles interact with the standard model particles via gravity and via the
kinetic mixing interaction \cite{holdom,fhe}, which also leads to photon - mirror photon kinetic mixing:
\begin{eqnarray}
{\cal L}_{\rm mix} = \frac{\epsilon}{2}  F^{\mu \nu} F^{'}_{\mu \nu}
\label{mix1}
\end{eqnarray}
where $F^{\mu \nu}$ [$F^{'}_{\mu \nu}$] is the field strength tensor of the photon 
[mirror photon].
The kinetic mixing interaction induces tiny ordinary electric charges for the charged
mirror sector particles of $\pm \epsilon e$ for the mirror proton and mirror electron respectively.

The dark matter particles, the mirror protons and mirror electrons, constitute the inferred dark matter in the Universe 
in this picture, e.g. \cite{review}.
The unbroken $Z_2$ mirror symmetry implies that the masses of the 
mirror proton and mirror electron are exactly identical to their ordinary matter counterparts.
More generally though, one can consider models with more generic hidden sectors, featuring dark electrons
and dark protons coupling together via massless dark photons, e.g. \cite{sunny}. In such models
the dark electron and dark proton masses are independent parameters. The analysis in this paper can be applied
to both mirror dark matter as well as more generic dark sector models.

%The dark matter particles, the dark protons and dark electrons, constitute the inferred dark matter in the Universe.
%Considerations from halo dynamics as well as other
%small scale structure issues  indicate that
%a kinetic mixing strength of around $\epsilon \approx 2\times 10^{-10}$ is favoured for mirror dark matter parameter space.
%The reader is referred to recent papers \cite{sny2,foot18a,foot18} and the review \cite{review} for further details.
%Here, the focus is on examining some aspects of the expected electron scattering signal in direct detection 
%experiments in such a model.

\vskip 0.3cm

\section{The electron scattering rate}

Since dark electrons are electrically charged in the presence of kinetic mixing, they can scatter off ordinary electrons. 
As mentioned earlier, keV energy recoils are kinematically possible due to energy
equipartition between the 
light dark electron and heavy dark nuclei halo
components.
Coulomb scattering of a dark electron off an electron is a spin-independent process.  
Approximating the target electron as free and at rest relative to the incoming dark
electron of speed $v$,
the cross section is
\begin{eqnarray}
\frac{d\sigma}{dE_R} = \frac{\lambda}{E_R^2 v^2} 
\label{cs}
\end{eqnarray}
where
$\lambda \equiv 2\pi \epsilon^2 \alpha^2/m_e$,
and $E_R$ is the recoil energy of the scattered electron. 
Naturally, treating the target electrons as free can only be approximately valid for the loosely bound atomic electrons, 
i.e. those with binding energy much less than $E_R$.
We define $g_T(E_R)$ as the number of electrons per target atom with atomic binding energy ($E_B$) less than $E_R$, and we approximate
the electron scattering rate per target atom by replacing $\lambda \to g_T \lambda$ in Eq.(\ref{cs}).
[For the DAMA experiment, the `atom' is a NaI pair.]
Typically, the proportion of loosely bound electrons, i.e. with $E_B \ll E_R$, greatly outnumbers those with $E_B \sim E_R$, so
this approximation is expected to be reasonable.

The scattering rate of a dark electron off an electron is then:
\begin{eqnarray}
\frac{dR_e}{dE_R} &=& 
g_T N_T  n_{e'} 
\int \frac{d\sigma}{dE_R}
\ f({\textbf{v}}; {\textbf{v}}_E; \theta)\ |{\textbf{v}}| \ 
d^3v \nonumber \\
&=& g_T N_T  n_{e'}
\frac{\lambda}{E_R^2} 
\ I({\textbf{v}}_E,\theta)
\label{55}
\end{eqnarray}
where
\begin{eqnarray}
I({\textbf{v}}_E,\theta)
\equiv \int^{\infty}_{|{\bf{v}}| > v_{min} (E_R)} 
\ \frac{f({\textbf{v}}; {\textbf{v}}_E; \theta)}{|{\textbf{v}}|} \ d^3 v
\ .
\label{III}
\end{eqnarray}
Here, $N_T$ is the number of target atoms per kg of detector,  
$v_{\rm min} \ = \ \sqrt{E_R m_e/2\mu^2}$, where $\mu$ is the electron - dark electron reduced mass ($\mu = m_e/2$ for the mirror
dark matter case),
%$v_{\rm min} \ = \ \sqrt{2E_R/m_e}$, 
and 
$n_{e'}$ is the dark electron number density at the detector's location. 
Also, $f({\textbf{v}}; {\textbf{v}}_E; \theta)$
is the velocity distribution of dark electrons which arrive at the detector. 
The detector is in motion, due to the daily rotation of the Earth, described in terms of the angle $\theta (t)$ 
to be defined shortly, and the Earth
itself is in motion around the Sun, with velocity ${\textbf{v}}_E$ of magnitude:
\begin{eqnarray}
v_E = v_\odot + \Delta v_E \cos \omega (t-t_0)\ .
\end{eqnarray}
Here, $\omega =  2\pi/{\rm year}$,
$v_\odot \approx v_{\rm rot} + 12$ km/s
(the  12  km/s  correction  is  due  to  the
Sun's peculiar velocity) and $\Delta v_E = 
15$  km/s results from the Earth's orbital motion.
Evidently, $v_E$ varies by $\pm \Delta v_E$ during the year with a 
maximum at $t = t_0 \simeq 153$ days (June $2^{nd}$).

Far from the Earth, the dark electron 
distribution,
$f({\textbf{v}}; {\textbf{v}}_E)$, 
might possibly be approximately Maxwellian,
%would be  very small. This follows because of the large
%$e'$ velocity dispersion, $v_0 = 11200\sqrt{T ({\rm keV})/0.35}$ km/s, given the halo temperature expected from Eq.(\ref{1}).
however 
%the distribution $f({\textbf{v}}; {\textbf{v}}_E)$ 
near the Earth it will be strongly influenced by the halo interactions with Earth-bound dark matter.
The Earth-bound dark matter forms an obstacle to the halo wind, which is moving through the halo
at roughly the speed of sound.
It turns out that even small changes to this speed, due to the Earth's motion around the Sun,
can lead to large effects for the halo dark matter  density
and distribution
near the Earth \cite{JC1}.
The effects of the halo interaction with captured dark matter within the Earth can thereby lead to a
strongly time varying distribution at the detector's location, $f({\textbf{v}}; {\textbf{v}}_E; \theta)$.
Not only is the distribution time varying, it would not be Maxwellian.
On the particle level, the distribution is strongly influenced by dark electromagnetic fields induced
in the halo plasma and in the Earth-bound dark matter, the outer layers of which can form a `dark ionosphere'.  
The dark electron distribution can also be influenced by collisions with
Earth-bound dark matter, as these interactions can effectively shield the detector from the halo wind.
%Significant anisotropy in the distribution might arise, as well as other major deviations. 
%Presumably  the anisotropy is not of critical importance for the DAMA experiment, since directional 
%information is not utilized.  

A simple model arises if the dark electron distribution has a mean speed, $\langle |\textbf{v}| \rangle$, 
which is much greater than $v_{\rm min}$.
% (for $E_R \stackrel{<}{\sim} {\rm few}$ keV). 
In that limit, $I({\textbf{v}}_E,\theta)$ becomes
approximately independent of $v_{\rm min}$, and consequently also independent of $E_R$.
Since $v_{\rm min} \propto \sqrt{E_R}$, the condition
$\langle |\textbf{v}| \rangle \gg v_{\rm min}$ 
could only be valid for electron recoil energies below some threshold, $E_R^T$.
For recoil energies greater than $E_R^T$ the electron scattering
rate becomes strongly suppressed, and falls much faster than $1/E_R^2$.
To have some possibility of explaining the DAMA signal, we will need $E_R^T \gtrsim 2-3$ keV. 

For mirror dark matter, the temperature of the dark halo is estimated to be around $T \approx 0.5$ keV, which at first glance would
suggest that the condition 
$\langle |\textbf{v}| \rangle \gg v_{\rm min}$ 
is unlikely to be valid for keV energy electron recoils.
This reasoning assumes an undistorted halo mirror electron energy spectrum, such as a Maxwellian distribution.
However, the effects of the Earth bound dark matter can strongly influence the halo mirror electron distribution that
arrives at the detector;
there might well be an
effective cutoff at low velocities due to the induced dark electromagnetic fields in the Earth's vicinity, or due to collisional shielding
of the halo wind by Earth-bound dark matter.
In more generic dark sector models there is more freedom, and the
halo temperature, Eq.(\ref{1}), can be relatively high, e.g. in the multi-keV range.
In either case, it seems possible that
the condition
$\langle |\textbf{v}| \rangle \gg v_{\rm min}$ 
might be roughly valid for keV energy electron recoils.

Independently of the details of the underlying model we can explore the time dependence of the velocity function,
$I({\textbf{v}}_E,\theta)$, via a Taylor series 
expansion,\footnote{Here, we treat $n_{e'}$ as fixed, so any time dependence of the number density of halo dark electrons in the vicinity of the detector
is also absorbed into 
the $I(\textbf{v}_E,\theta)$ Taylor series expansion.}
%keeping only the linear terms,
\begin{eqnarray}
I({\textbf{v}}_E,\theta) = I_0  + \frac{\partial I}{\partial v_E} \Delta v_E \cos \omega (t-t_0) + \frac{\partial I}{\partial \theta} (\theta - \bar \theta)
+ ...
\ .
\label{t11}
\end{eqnarray}
Here, $\theta(t)$ is the angle between the direction of the halo wind
and the zenith at the detector's location. 
At the Gran Sasso latitude
this angle has a large daily variation as well as a small annual modulation
($\bar \theta \simeq 2.17$ is the average at Gran Sasso). 
See Ref.\cite{JC1} for further discussion.
As discussed above, 
$I({\textbf{v}}_E,\theta)$ becomes independent of $E_R$ for sufficiently low recoil energies, $E_R < E_R^T$, and thus
the coefficients $I_0$, $\partial I/\partial v_E$, $\partial I/\partial \theta$ are also $E_R$ independent.

The net result is a rather simple phenomenological model with a time varying electron scattering rate: 
\begin{eqnarray}
\frac{dR_e}{dE_R} = 
g_T N_T n_{e'} \frac{\lambda}{v_c^0 E_R^2}\left[ 1 + A_v\cos\omega (t-t_0) + A_\theta (\theta - \bar \theta)\right]
\ \ \ {\rm for} \ E_R \lesssim E_R^T
%\ \  {\rm for} v_{\rm min} < v_c^0 
%\ .
\label{r68}
\end{eqnarray}
where $v_c^0 \equiv 1/I_0$ and $E_R^T \sim 2\mu^2 (v_c^0)^2/m_e$.
For $E_R \lesssim E_R^T$, an approximate $dR_e/dE_R \propto 1/E_R^2$ behaviour for both the average rate and the annual modulation is predicted
(putting aside the relatively minor $E_R$ dependence of $g_T$).
For $E_R \gtrsim E_R^T$,  the scattering rate becomes suppressed, falling off much faster than $1/E_R^2$.
It should be clear that Eq.(\ref{r68}) is applicable to mirror dark matter as well as more generic plasma dark matter models.

Finally, note that the parameter $A_\theta$ controls the diurnal variation. Published DAMA results indicate that this parameter
is consistent with zero \cite{damadiurnal}, although there is a hint of a diurnal signal at $2.3\sigma$ C.L \cite{foot14,JC1}.
We note that the DAMA phase 2 diurnal variation results have yet to be reported, and it will be interesting
to see if stronger evidence for a daily variation emerges.
For the purposes of this paper, though, we set $A_\theta = 0$.
%Our main focus here is on the annual modulation and averge rate.

\section{Implications for direct detection experiments}

With $A_\theta = 0$,
the electron scattering rate given in Eq.(\ref{r68}) is defined in terms of the parameters, 
$\epsilon \sqrt{n_{e'}/v_c^0}$, $A_v$, $E_R^T$.
%not known, For mirror dark matter, $n_{e'} \sim n_0 \equiv 0.2 \ {\rm cm^{-3}}$
For definiteness, we shall assume that $E_R^T \gtrsim 6$ keV so that Eq.(\ref{r68})
is valid for the entire DAMA region of interest. A somewhat lower $E_R^T$ is, of course, still possible, but would need
to cover at least part of the DAMA energy region of interest (i.e. $E_R^T \gtrsim 2-3$ keV). 
The remaining parameters,  
$\epsilon \sqrt{n_{e'}/v_c^0}$, $A_v$.
will be constrained by comparing the
electron scattering rate with the data from the most relevant experiments, including the 
DAMA annual modulation signal.

To obtain the predicted rate for a given experiment, 
the detection efficiency and energy resolution will need to be modelled. This is done by
convolving the rate:
\begin{eqnarray}
\frac{dR_e}{dE_R^{m}} = \int {\cal G}(E_R^m,E_R) \frac{dR_e}{dE_R} \epsilon_F(E_R) \ dE_R
\ .
\label{r4}
\end{eqnarray}
Here, ${\cal G}(E_R^m,E_R)$ is the resolution function taken to be a Gaussian, and $\epsilon_F$ is the detection efficiency.

% with 
%the resolution given as measured in \cite{damares}, while $\epsilon(E_R)$ is the detection
%efficiency which we set to unity as the DAMA collaboration gives their results corrected for detection efficiency.

\subsection{DarkSide-50}

We first consider constraints on the time-average electron scattering rate, and we then consider the DAMA annual modulation
signal. 
The strongest constraint on the average rate arises from the DarkSide-50 experiment \cite{darkside}.
The limits follow from
an analysis of the ionization signal with a 6786.0 kg-day exposure of an Argon target.
The DarkSide experiment features an analysis threshold of 0.05 keV. However since the rate below 0.1 keV
is believed to include new sources of unmodelled background, including coincident single electron events, as described in \cite{darkside,darkside1}, 
we prefer the 0.1 keV threshold, also used in \cite{darkside1}. 
The low threshold of the DarkSide experiment makes it particularly sensitive to the dark matter electron scattering 
due to the sharply increasing rate at low energies: $dR_e/dE_R \propto 1/E_R^2$.

To compare the electron scattering rate, Eq.(\ref{r68}), with the DarkSide data, we 
take into account the energy resolution using $\sigma/E_R = 0.5$ \cite{pc}, and efficiency $\epsilon_F = 0.43$ \cite{darkside1}.
The DarkSide data are given in terms of the number of extracted electrons, $N_{e}$,
which, in the low energy region of interest ($4 \le N_e \le 15$),
is related to the recoil energy of the scattering event via $N_{e} \approx 40 E_R^m/{\rm keV}$.
The expected rate in their experiment is then $dR_e/dN_e = [dR_e/dE_R^m] [dE_R^m/dN_e]$.
%we can derive a rough upper limit on $\epsilon \sqrt{n_{e'}/n_0}/\sqrt{v_c^0}$. 
An excess above modelled backgrounds is present in the DarkSide data near their low energy threshold. 
If we tentatively
assign this excess to electron scattering, 
we obtain an estimate of
$\epsilon \sqrt{n_{e'}/v_c^0}$, equivalent to:
\begin{eqnarray}
\epsilon \sqrt{\frac{n_{e'}}{0.2 \ {\rm cm^{-3}}}} 
\approx 1.5 \times 10^{-11}
\sqrt{\frac{v_c^0}{50000 \ {\rm km/s}}} 
\ .
\label{fix}
\end{eqnarray}
The modelled rate, along with the DarkSide data, is shown in Figure 1.
The figure indicates that the observed low energy excess near the DarkSide threshold 
is compatible the $1/E_R^2$ scaling predicted by this model.

\begin{figure}[t]
    \centering
    \includegraphics[width=0.46\linewidth,angle=270]{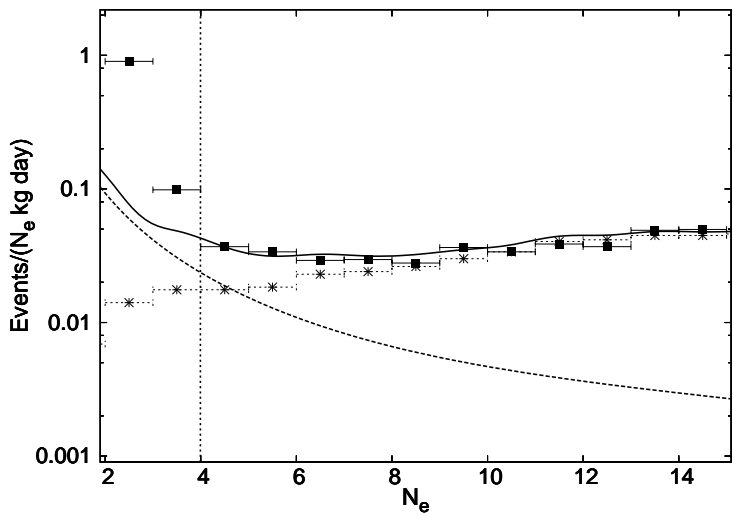}
    \vspace{4ex}
    \vskip -1.0cm
\caption{
\small
The DarkSide ionization spectrum (solid error bars)
compared with the modelled electron recoils from dark electron interactions + background (solid curve).
Also shown are the DarkSide estimated background rate (dotted error bars) and the predicted dark matter scattering rate    
(dashed curve). The threshold at $N_e =4$ is also indicated.
}
\end{figure}

\subsection{XENON100-S1}

The XENON100 collaboration have searched for low energy electron recoils in the (2-6) keV region 
and also obtained fairly tight constraints on the dark matter electron scattering rate \cite{xene2}.
%In addition they have searched for an annual modulation obtaining a slight
%average annual modulation of around 0.002 cpd/kg/keV in the (2-6) keV region.
With the 
$\epsilon \sqrt{n_{e'}/v_c^0}$ parameter 
given in Eq.(\ref{fix}), we can compare
the expected rate with the XENON100 data. 

The XENON100 analysis uses the prompt photon signal (S1).
At around 2 keV,
the S1 signal (unlike S2) falls sharply due to detection efficiency, 
and other effects.
To model this sharp feature, we use a low energy effective cutoff at $E_R = 2.2$ keV.
That is, we take the detection efficiency function as $\epsilon_F = 1$ for $E_R > 2.2$ keV and zero
otherwise.
For the energy resolution,
the XENON100 paper indicates that it is twice worse than that of DAMA, which
roughly corresponds to, $\sigma/E_R = 0.9/\sqrt{E_R/{\rm keV}}$.
The XENON100 data are given in terms of the number of S1 photoelectrons ($N_{\rm PE}$).
The conversion between detected electron recoil energy and $N_{\rm PE}$ is roughly, $N_{\rm PE} = 5[E_R^m/(3\ {\rm keV})]^{1.4}$, so that
$N_{\rm PE} = 3$ corresponds to $E_R^m \approx 2$ keV.
The modelled rate, along with the XENON100 S1 data, are shown in Figure 2.

\begin{figure}[t]
    \centering
    \includegraphics[width=0.46\linewidth,angle=270]{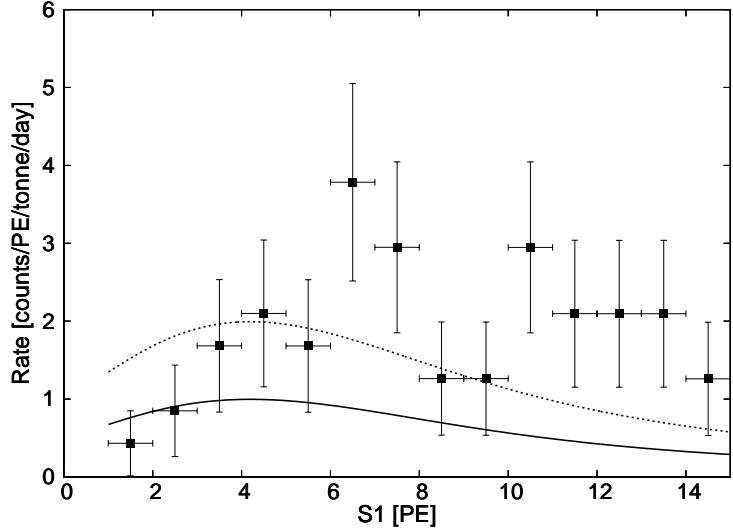}
    \vspace{4ex}
\vskip -0.8cm
\caption{
\small
The XENON100 electron recoil data collected during 70 days near the expected yearly maximum on June $2^{nd}$ \cite{xene2}. 
Also shown are the yearly average electron recoil rate expected from dark electron interactions (solid line)
for the same
$\epsilon \sqrt{n_{e'}/v_c^0}$ parameter as per Figure 1 [Eq.(\ref{fix})].
The dotted line indicates the
maximal rate assuming maximal annual modulation, $A_v = 1$.
}
\end{figure}

The XENON100 S1 data were collected during 70 days near the expected yearly maximum on June $2^{nd}$,
so the predicted rate can be up to a factor of two larger than the yearly average rate during this period if the annual modulation is maximal.
Even with such an enhancement,
Figure 2 indicates that the electron scattering rate is  roughly consistent with the XENON100 S1 data (which
additionally contains an uncertain unmodelled background component).
The first bin appears somewhat low, but this might not be very significant in view of the uncertainties 
in modelling the energy resolution, detection efficiency etc.

We have also examined XENON100 S2 data from \cite{s2}. However, the estimated rate
in that experiment turns out to be more than an order of magnitude below their observed rate,
and therefore does not pose any constraint on this model.

%In addition they have searched for an annual modulation obtaining a slight
%average annual modulation of around 0.002 cpd/kg/keV in the (2-6) keV region.

\subsection{The DAMA annual modulation}

We now consider the DAMA experiment. The DAMA collaboration have measured an annually modulated scintillation rate using
a NaI target
in their low energy region (1-4 keV), with phase consistent with dark matter interactions.
In the dark matter model discussed here, the annual modulation is set by the parameter $A_v$ in Eq.(\ref{r68}).
By construction, $A_v \le 1$, with $A_v = 1$ corresponding to maximal annual modulation.

To evaluate the predicted rate we use the measured
DAMA resolution  of $\sigma/E_R = 0.448/\sqrt{E_R/{\rm keV}} + 0.009$ \cite{damares}, and set the detection
efficiency to unity as the DAMA collaboration give their results corrected for detection efficiency.
This can only be a rough approximation, and a more sophisticated analysis should find a softening of the annual
modulation in the low energy region due to the falling efficiency below the threshold energy.

We have evaluated the predicted electron scattering rate, again
fixing the $\epsilon \sqrt{n_{e'}/v_c^0}$ parameter by Eq.(\ref{fix}).
The predicted annual modulation amplitude
is proportional to $A_v$, and the
result for maximal annual modulation, $A_v = 1$, is shown in Figure 3.
The predicted annual modulation is somewhat less than the observed rate for $2 < E_R/{\rm keV} <  4$, although there are
potentially substantial uncertainties
(especially in modelling the energy resolution and detection efficiency).
If $A_v$ is substantially less than unity, or if the DarkSide-50 low energy excess is due to some unmodelled background component (rather
than dark matter interactions) then the predicted annual modulation amplitude for the DAMA
experiment would be much lower, presumably too low to possibly account for the DAMA signal.
To have a chance of explaining the observed annual modulation rate in DAMA, it therefore seems likely that
the annual modulation is near maximal, $A_v \approx 1$, and  that the DarkSide-50 low energy excess is due
primarily to dark matter interactions.
An annual modulation analysis of the DarkSide-50 data should provide a useful test of this logic.

\begin{figure}[t]
    \centering
    \includegraphics[width=0.46\linewidth,angle=270]{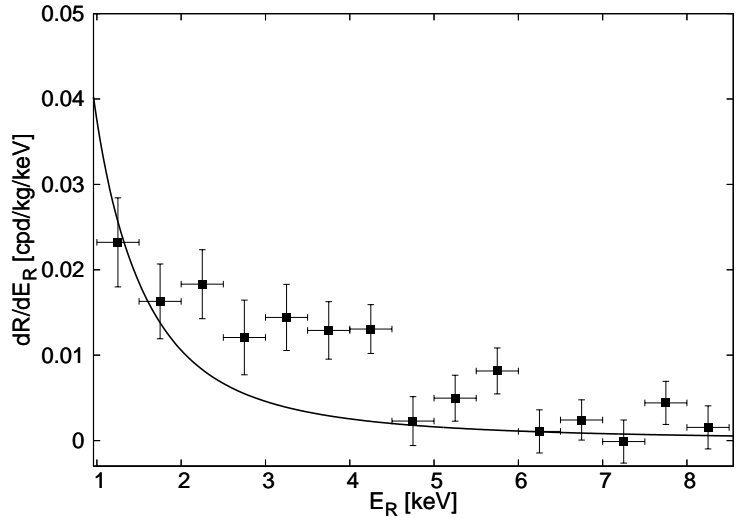}
    \vspace{4ex}
   \vskip -0.9cm
\caption{
\small
The annual modulation amplitude predicted for the DAMA experiment, with $A_v = 1$ (maximal).
The data is from \cite{dama4}.
The $\epsilon \sqrt{n_{e'}/v_c^0}$ parameter adopted [Eq.(\ref{fix})]  is the same as per Figures 1,2. 
}
\end{figure}

The XENON100 collaboration also searched for an annual modulation in the (2-6) keV region with
the aim of testing models explaining the DAMA  
signal via electron scattering \cite{xene1}. They obtained some interesting results, including a hint of an annual modulation, close to maximal,
with the same phase as DAMA. 
%On the although the annual modulation amplitude was constrained to be rather low.
(A fit of the XENON100 data to an annually modulating rate found a nonzero amplitude at 
$2.4 \sigma$ C.L., with best fit amplitude around 2 cpd/tonne/keV.)
For the same parameters as per Figs.1-3 [i.e. $A_v = 1$, and $\epsilon \sqrt{n_{e'}/v_c^0}$ given by Eq.(\ref{r68})], we 
find that this model predicts an annual modulation amplitude for the 
XENON100 experimental setup (S1 signal) of 1.8 cpd/tonne/keV in the
(2-6) keV region. 
This is close to the best fit value identified in the XENON100 annual modulation analysis.

To summarize, both the average rate of XENON100 electron recoils and their observed time dependence,
seem to be compatible (allowing for reasonable uncertainties) with 
this dark electron scattering interpretation of the DAMA annual modulation signal.
%From the analysis above, it therefore seems possible that
%the small annual modulation rate observed by XENON100 is not inconsistent with that found in the DAMA experiment.
The basic reason for the relatively low average rate and modulation amplitude of electron recoils in the XENON100 experiment can be qualitatively
understood as follows:
Given the sharply rising scattering rate at low energy,
$dR_e/dE_R \propto 1/E_R^2$, 
much of the DAMA annual modulation signal is due to recoils
with actual energy below their threshold, detectable only because of the poor energy resolution.
(Naturally, this energy resolution might well need to deviate from Gaussian behaviour in the tail to reproduce the
observed modulation above 2 keV).
By contrast,
in the XENON100 experimental setup, recoils with actual energy below around 2 keV might not be observable due to the
sharply falling XENON100 S1 signal efficiency near 2 keV.
The analysis of the XENON100 collaboration, as given in \cite{xene1,xene2}, does not take
into account this effect when they directly compare their data with that from the DAMA experiment in the 2-6 keV energy range.

\section{Conclusion and Discussion}

Within the mirror dark matter model, and some related models, the dark halo of the Milky Way is expected
to take the form of a dark plasma.
In the mirror dark matter case,
this plasma consists of light mirror electrons of mass $m_e \simeq 0.511$ MeV
and heavier mirror ions, including mirror protons, mirror helium nuclei, and possibly
heavier mirror metal components.
Provided that the kinetic mixing interaction exists,
the mirror electrons and mirror ions in the plasma can potentially produce keV electron
recoils and nuclear recoils respectively.
It has been argued previously \cite{foot14,JC1} that mirror electron scattering off electrons might
be responsible for the DAMA annual modulation signal, especially as such an electron scattering
interpretation is relatively weakly constrained by other experiments.
In light of the new results from DAMA \cite{dama4}, we have reconsidered this interpretation.

It is difficult to estimate
the rate of electron recoils in a direct detection experiment due to the
inherent complexities involved in modelling the halo plasma interactions with Earth-bound dark matter.
Here, we have considered a simple phenomenological model for the dark matter velocity function 
relevant for direct detection experiments. 
This model predicts an approximate 
$dR_e/dE_R \propto 1/E_R^2$ 
behaviour for both the average scattering rate and the annual modulation.
Such a recoil spectrum is roughly compatible with the annual modulation
as measured in the DAMA experiment.
The average rate is consistent with the results from other experiments, with the DarkSide experiment providing
the most useful information.
That experiment sees an excess of events at low energies which can be interpreted as dark matter induced electron 
recoils, and
in combination with DAMA, indicates that the annual modulation amplitude is near maximal.
This conclusion is supported by XENON100 results as analysed here. 

%The modulation amplitude was found to be typically very large, approaching maximal for some parameter values.
%As a consequence, the average rate is quite low, but still has some tension with 
%the constraints reported by XENON100 \cite{xene1,xene2}.

%With regard to the time dependence of the scattering rate, it is in general not sinusoidal, although
%it can be approximately sinusoidal at high recoil energies.
%At lower energies, there is a transition and the rate deviates from 
%sinusoidal behaviour below around, 1-3 keV. 
%Such distortions might ultimately prove useful in testing
%the model.

Although the model is phenomenological to some extent,
this explanation of the DAMA signal can be tested in the near future. 
For example, if XENON1T shows an
electron scattering rate below $0.001$ cpd/kg/keV at around 2 keV,  then this explanation will be disfavoured.
If they see a rising event rate, a 
$dR_e/dE_R \propto 1/E_R^2$ behaviour is anticipated, as described above.
An analysis of the DarkSide low energy excess between $4 \le N_E \le 8$ should reveal
a significant annual modulation with the same phase as the DAMA signal.
A more critical
test of the model resides in the predicted sidereal daily modulation.
Previous published results of DAMA \cite{damadiurnal} already contained a $2.3\ \sigma$ hint of such a variation with 
the correct phase \cite{foot14,JC1}.
It is likely that experiments at lower latitudes, including XMASS and PANDA, could have a larger diurnal
variation, and we encourage these experiments to give results for this search channel.
For an experiment located in the southern hemisphere, including the proposed SABRE experiment \cite{sabre}, 
the diurnal modulation can be maximal and provide an even more rigorous test of the ideas discussed here.

%In fact, such a detector could discover dark matter in just a few months of operation.

%That is, a manifestation of the differing response of these detectors to low energy recoils.

\vskip 1.2cm
\noindent
{\large \bf Acknowledgments}

\vskip 0.3cm
\noindent
The author would like to thank J. Collar and R. Lang for helpful correspondence, and 
especially M. Szydagis
for invaluable assistance.
The referee is also thanked for suggesting valuable improvements to the paper.
This work was supported by the Australian Research Council.

%\newpage

\end{document}